\documentclass[%
 reprint,
superscriptaddress,
 amsmath,amssymb,
 aps,
pre,
floatfix,
]{revtex4-2}
\usepackage{relsize}
\usepackage{graphicx}
\usepackage{xfrac}    

\usepackage{physics}
\usepackage{dcolumn}
\usepackage{bbm}
\usepackage[dvipsnames]{xcolor}
\usepackage[colorlinks=true,citecolor=teal, urlcolor=Blue, linkcolor=YellowOrange]{hyperref}
\usepackage{mathtools}

\newcommand{\sset}[1]{\left\{ #1 \right\}} 

\newcommand{\virg}[1]{``#1''}

\usepackage[capitalise]{cleveref}

\usepackage{pdfpages} 
\usepackage{pgffor} 

\makeatletter
\AtBeginDocument{\let\LS@rot\@undefined}
\makeatother

\def\supplementfilename{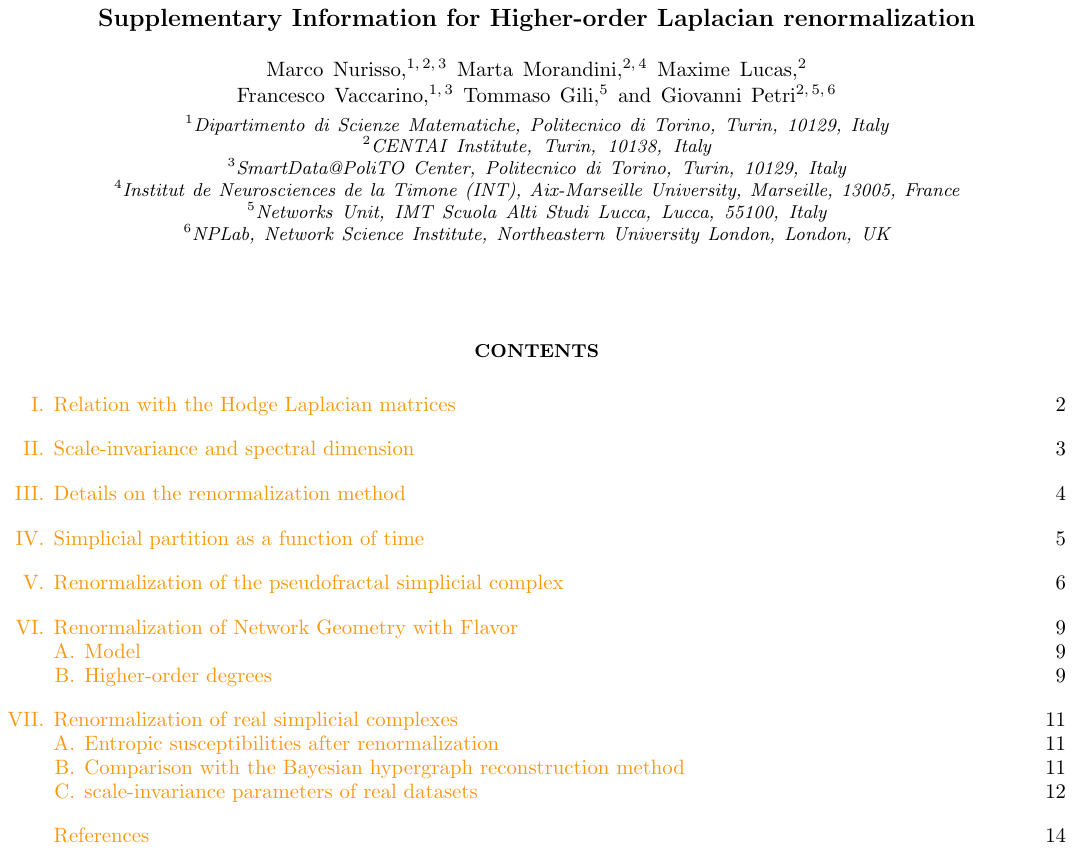}

\pdfximage{\supplementfilename}
\def\numbersupplementpages{\the\pdflastximagepages}

\begin{document}


\title{Higher-order Laplacian Renormalization}

\author{Marco Nurisso}
 \affiliation{Dipartimento di Scienze Matematiche, Politecnico di Torino, Turin, 10129, Italy}
 \affiliation{CENTAI Institute, Turin, 10138, Italy}
 \affiliation{SmartData@PoliTO Center, Politecnico di Torino, Turin, 10129, Italy}
\author{Marta Morandini}%
\affiliation{CENTAI Institute, Turin, 10138, Italy}
\affiliation{Institut de Neurosciences de la Timone (INT), Aix-Marseille University, Marseille, 13005, France}

\author{Maxime Lucas}
\affiliation{CENTAI Institute, Turin, 10138, Italy}

\author{Francesco Vaccarino}
\affiliation{Dipartimento di Scienze Matematiche, Politecnico di Torino, Turin, 10129, Italy}
\affiliation{SmartData@PoliTO Center, Politecnico di Torino, Turin, 10129, Italy}

\author{Tommaso Gili}
\email{tommaso.gili@imtlucca.it}
\affiliation{Networks Unit, IMT Scuola Alti Studi Lucca, Lucca, 55100, Italy}
\author{Giovanni Petri}
\email{giovanni.petri@nulondon.ac.uk}
\affiliation{NPLab, Network Science Institute, Northeastern University London, London, UK}
\affiliation{CENTAI Institute, Turin, 10138, Italy}
\affiliation{Networks Unit, IMT Scuola Alti Studi Lucca, Lucca, 55100, Italy}

\begin{abstract}
We propose a cross-order Laplacian renormalization group (X-LRG) scheme for arbitrary higher-order networks.
The renormalization group is a pillar of the theory of scaling, scale-invariance, and universality in physics. 
An RG scheme based on diffusion dynamics was recently introduced for complex networks with dyadic interactions. 
Despite mounting evidence of the importance of polyadic interactions, we still lack a general RG scheme for higher-order networks. 
Our approach uses a diffusion process to group nodes or simplices, where information can flow between nodes and between simplices (higher-order interactions).
This approach allows us (i) to probe higher-order structures, defining scale-invariance at various orders, and (ii) to propose a coarse-graining scheme.
We demonstrate our approach on controlled synthetic higher-order systems and then use it to detect the presence of order-specific scale-invariant profiles of real-world complex systems from multiple domains.
\end{abstract}

\maketitle

\section{Introduction}

The renormalization group (RG) \cite{fisher1974renormalization} is a cornerstone of modern theoretical physics because it allows us to study how a physical system depends on the scale of observation, defining universality classes and, importantly, formalizing the concept of scale-invariance.
While the RG has been a powerful tool for understanding a broad class of physical systems, extending its framework to complex networks has posed a recent and significant challenge, mainly due to the correlations between scales caused by small-world effects \cite{tu2023renormalization}.
This challenge has gained substantial attention \cite{song2005self,goh2006skeleton,rozenfeld2010small,garcia2018multiscale,garuccio2020multiscale,zheng2023geometric,Loures2023Feb,villegas2023laplacian} due to its potential to provide insights into the multiscale structural organization of complex networks.

Notable approaches \cite{serrano2008self,garcia2018multiscale} are based on the hypothesis that an embedding space exists and is responsible for the network structure.
This underlying geometry provides a natural way to identify groups of nearby nodes, reminiscent of spin blocks in the traditional real-space RG process \cite{kadanoff1966scaling}. 
These groups can then be collapsed into ``super-nodes'', providing a coarse-grained network description.
However, this perspective encounters a fundamental limitation: networks are inherently topological structures, devoid of geometry, and thus need a topological notion of RG \cite{garuccio2020multiscale,boguna2021network}.

Diffusion provides such a notion: it is a dynamical process that one can define on any combinatorial structure and depends only on the structure's topology.
Diffusion on graphs describes the dynamics of information flowing from node to node through edges, eventually becoming uniformly distributed on its connected components.
This process is formalized as a first-order system of linear differential equations specified by the graph Laplacian matrix $\mathbf{L}$. 
One can then see the diffusion time as a \textit{resolution parameter}: at short times, information only diffuses to neighboring nodes, revealing local structure; at longer times, diffusion reaches nodes further apart and reveals the global network structure.
A recent proposal, the Laplacian renormalization group (LRG) scheme \cite{villegas2023laplacian}, leverages this observation to produce coarser descriptions of a network's structure by identifying groups of nodes that are strongly linked by diffusion at a given scale.
Moreover, one can adopt this same approach to define an \emph{informational} notion of \emph{scale-invariance}, based on the properties of the diffusion process via the Laplacian spectrum \cite{bianconi2020spectral}, and thus different from the canonical concept of scale-freeness, which instead depends directly on the degree distribution.
Both of these results hinge on the formalism of network density matrices \cite{de2016spectral} to describe the complete behavior of information diffusion at a given scale.

Networks, however, are only part of the story. 
Great attention has recently been devoted to the study of higher-order networks: networks that encode multi-node interactions, going beyond the pairwise interactions of traditional networks \cite{battiston2020networks,battiston2021physics,bick2023higher}.
Higher-order interactions are present in many natural systems and drastically impact most dynamical processes, such as random walks \cite{schaub2020random,mukherjee2016random,parzanchevski2017simplicial}, diffusion \cite{muhammad2006control,torres2020simplicial}, spreading \cite{lucas2023simplicially,simplicial_contagion,chowdhary2021simplicial,st2022influential, ferraz2023multistability,chen2024simplicial,kiss2023insights}, coordination \cite{alvarez2021evolutionary,iacopini2023temporal,mancastroppa2023hyper,shang2022system,neuhauser2020multibody,neuhauser2021consensus,sahasrabuddhe2021modelling}, and synchronization \cite{skardal2019abrupt,lucas2020multiorder,zhang2023higher,gambuzza2021stability,millan2020explosive,nurisso2023unified}.
A higher-order interaction between $k+1$ nodes is typically called a simplex of order $k$ (or equivalently $(k+1)$-hyperedges). 
Systems with such interactions are formalized using simplicial complexes or hypergraphs, with the former being more structured, as the presence of interaction also requires the presence of interactions between all of its node subsets. 
  
However, little work exists on RG approaches to higher-order networks despite their importance. 
A direct generalization of the Laplacian RG approach \cite{villegas2023laplacian}, based on the multiorder Laplacian \cite{lucas2020multiorder}, was recently proposed \cite{cheng2023simplex}. 
This proposal is, however, \textit{node-centric}: it only considers the diffusion of information from node to node.
A parallel research line \cite{bianconi2020spectral, reitz2020higher, sun2020renormalization} focused on specific families of simplicial complexes and used renormalization group techniques to compute some notable statistical properties.

Here, we propose a general renormalization group scheme for arbitrary higher-order networks. 
Our approach uses a higher-order notion of diffusion that we formally define by introducing the \textit{cross-order Laplacian}.
In this new diffusion process, information can flow between simplices of any order $k$ via simplices of any other order $m$.  
This proposal provides a natural generalization of previous ones \cite{villegas2022laplacian,cheng2023simplex}, which are restricted to node-node diffusion.
By studying the properties of this diffusion via the cross-order Laplacians, our approach allows us to probe the existence of characteristic scales, or---crucially---their absence (\textit{scale-invariance}) in higher-order networks at each order.
In particular, we first define the appropriate Laplacian matrices to describe generic higher-order diffusion. 
We then leverage them to define (i) a higher-order notion of informational scale-invariance through the von Neumann entropy and entropic susceptibility, and (ii) an explicit RG scheme informed by a chosen higher-order diffusion process. 
Using these tools, we extract a cross-order scale signature in simplicial complexes obtained from synthetic models and real-world data and show that in most cases, scale-invariance is found only under the lens of specific orders, suggesting the existence of underlying order-specific processes. 

\begin{figure*}
    \centering
    \includegraphics[width = 0.8 \linewidth]{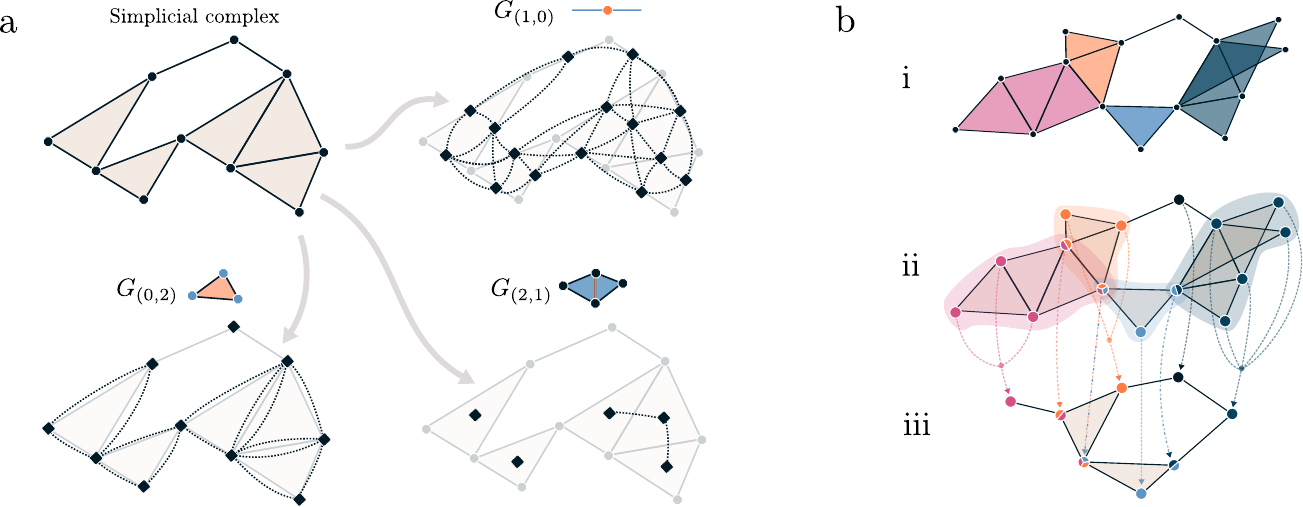}
    \caption{\textbf{Cross-order Laplacian renormalization scheme: partition and coarse-grain.} {\bf a.} Schematic representation of how the adjacency graphs are built for a simplicial complex. Notice how there can be cases like $G_{(0,2)}$, where the edges are weighted (here represented with multi-edges) because two vertices can be connected by more than one triangle. {\bf b.} Pictorial representation of our higher-order coarse-graining scheme with $k = 2$: (i) a partition of the $2$-simplices of a simplicial complex, here represented in color, is obtained through $\zeta$, (ii) each vertex inherits a signature containing the labels of all $2$-simplices it belongs to, (iii) vertices with the same signature are glued together and simplices are induced from the starting simplicial complex. }
    \label{fig:adjacency_graph}
\end{figure*}
\section{Higher-order networks and their structure}\label{section:structure_and_laplacians}
\paragraph{\textbf{Higher-order networks.}}
Let $\Delta$ be a higher-order network (also named hypergraph) on a finite set of vertices $V$, i.e., a family of subsets of $V$. 
Any element of $\Delta$ is called \emph{simplex} (or hyperedge) and its \emph{order} is defined as its cardinality reduced by one. 
A vertex will thus be a $0$-simplex, an edge a $1$-simplex, a triangle a $2$-simplex, and so on. 
If $\eta\in\Delta$ is a subset of a $k$-simplex $\sigma$, then we say that $\eta$ is a \emph{face} of $\sigma$.
We write $\Delta_k$ to denote the set of $k$-simplices of $\Delta$ and call $n_k$ its cardinality.
The object $\Delta$ is a simplicial complex when closed under inclusion, that is, $\sigma \in \Delta,\ \eta\subseteq\sigma \implies \eta\in\Delta$. 

Having fixed an order $k\in\mathbb{N}$, we want to find whether the hypergraph possesses characteristic scales or is scale-invariant from the perspective of $k$-simplices. 
In Ref. \cite{villegas2023laplacian}, it is argued that a diffusion process can be seen as a telescopic scanner of a (pairwise) network capable of extracting multiscale information about its structure.
This fact suggests studying our hypergraph's $k$-th order properties through a diffusion process on the $k$-simplices. \\

\paragraph{\textbf{Cross-order Laplacians.}}
A diffusion-like process on the simplices of a higher-order network can be defined in multiple ways, each associated with a different Laplacian matrix. 
In the case of simplicial complexes, the $k$-th order combinatorial Hodge Laplacian, proposed by Eckmann~\cite{eckmann1944harmonische}, is the one most commonly considered~\cite{lim2020hodge,krishnagopal2021spectral}, mainly for its deep connections with topology~\cite{lim2020hodge}. 
A diffusion process on the $k$-simplices \cite{muhammad2006control} can indeed be defined with the Hodge Laplacian, making it a natural candidate for our approach. 
However, it is possible to see that diffusion through the Hodge Laplacian does not correspond to a ``standard'' diffusion process as, for instance, the total amount of information flowing between simplices is not conserved (more details can be found in the Supplementary Information Section 1).
Due to this lack of clear physical interpretability, it is difficult to directly employ it in the Laplacian renormalization framework proposed in Ref.~\cite{villegas2022laplacian}. 
Moreover, the Hodge Laplacian cannot be naively extended to the general hypergraph setting.

Thus, we define a new family of Laplacian matrices that can describe a plethora of higher-order relations while maintaining a form analogous to the canonical graph Laplacian.
We do this by taking inspiration from the general theory of combinatorial complexes~\cite{hajijtopological} and the hypergraph Laplacian~\cite{aktas2022hypergraph}. 

Formally, we fix a number $k\in\mathbb{N}$, which we call \emph{diffusion order}, to describe a diffusion process on the $k$-simplices of the hypergraph.
We then need to decide how the process occurs, specifically, how simplices are \virg{connected} so that information can flow between them. 
The most natural approach is to extract this information from the structure of $\Delta$ by employing a notion of \emph{adjacency} among simplices.
In general, two $k$-simplices $\sigma$ and $\eta$ can be adjacent in two ways:
\begin{itemize}
    \item $\sigma,\eta$ are \emph{$m$-adjacent from above}, when there is an $m$-simplex $\xi$, with $m>k$, containing both of them $\xi\supseteq\sigma\cup\eta$;
    \item $\sigma,\eta$ are \emph{$m$-adjacent from below}, when they share a common $m$-face $\xi \subseteq \sigma\cap\eta$.
\end{itemize}
Combining these two definitions, we can define the \emph{adjacency number} $a_{(k,m)}$ of a pair of \emph{distinct} $k$-simplices $\sigma\neq \eta$ as
\begin{equation}
a_{(k,m)}(\sigma,\eta) = \begin{cases}
|\sset{\lambda\in\Delta_m: \lambda\subseteq\sigma\cap\eta}| \text{ if } m < k\\
|\sset{\lambda\in\Delta_m: \sigma\cup\eta\subseteq\lambda}| \text{ if } m > k\\
0 \text{ if } m = k
\end{cases}
\end{equation}
so that we may consider $\eta$ and $\sigma$ to be $m$-adjacent $\sigma\overset{m}{\sim}\eta$ when $a_{(k,m)}(\sigma,\eta) > 0$.
Intuitively, $a_{(k,m)}(\sigma,\eta)$ counts the number of $m$-simplices connecting $\sigma$ and $\eta$, allowing us to differentiate between different \virg{strengths} of adjacency.
For example, $a_{(2,0)} = 1$ when two triangles share a single vertex and $a_{(2,0)} = 2$ when they share an edge. 

The $(k,m)$-adjacency relations can be formalized into different \emph{adjacency matrices}, analogous to those defined in Refs.~\cite{hajijtopological,estrada2018centralities}. 
If we index the $k$-simplices as $\sigma_1,\dots,\sigma_{n_k}$, we can define the $(k,m)$-adjacency matrix with \emph{diffusion order} $k$ and \emph{interaction order} $m$ as the square $n_k\times n_k$ matrix with elements
\begin{equation}
    (\mathbf{A}_{(k,m)})_{ij} = a_{(k,m)}(\sigma_i,\sigma_j).
\end{equation}
The adjacency matrix $\mathbf{A}_{(1,2)}$, for example, describes how edges ($1$-simplices) are connected through triangles ($2$-simplices), while $\mathbf{A}_{(3,0)}$ tells us how tetrahedra ($3$-simplices) are attached to one another through vertices ($0$-simplices).

The matrix $\mathbf{A}_{(k,m)}$, notice, can be seen as the adjacency matrix of a (weighted) graph, which we call \emph{adjacency graph} $G_{(k,m)}$ (see \Cref{fig:adjacency_graph}a), whose nodes are the $k$-simplices and the edges are their adjacency relations given by $m$-simplices. 
In particular, $G_{(0,1)}$ is the graph underlying the higher-order network (i.e. $\Delta_0\cup\Delta_1$) and $G_{(1,0)}$ corresponds to its line graph \cite{whitney1992congruent}.
To each weighted adjacency matrix, we can associate its (weighted) Laplacian, which we name \emph{cross-order Laplacian}, through the usual formula
\begin{equation}\label{eq:diffusion_laplacian}  
\mathbf{L}^{\times}_{(k,m)} = \mathrm{diag}(\mathrm{deg}_{(k,m)}) - \mathbf{A}_{(k,m)},
\end{equation}
where $\mathrm{deg}_{(k,m)}$ is the vector containing the higher-order $(k,m)$-degrees, defined as the row-sums of $\mathbf{A}_{(k,m)}$.
More specifically, 
\begin{equation}\label{eq:degree}
\mathrm{deg}_{(k,m)}({\sigma}) =  \sum_{\eta\in\Delta_k} a_{(k,m)}(\sigma,\eta).
\end{equation}

The matrix $\mathbf{L}^{\times}_{(k,m)}$, being a weighted graph Laplacian, is symmetric, positive semidefinite and has an eigenvalue $0$ with multiplicity given by the number of connected components of its underlying graph $G_{(k,m)}$. 
We note that our definition includes two existing families of Laplacians as particular cases: (i) the \virg{vertex-centric} higher-order Laplacians describing how vertices can exchange information through simplices (e.g., the generalized Laplacians of Ref.~\cite{lucas2020multiorder}), as cross-order Laplacians of the form $\mathbf{L}^{\times}_{(0,m)}$ and (ii) Hodge-like Laplacians where diffusion happens between simplices through simplices of a directly adjacent order, as $\mathbf{L}^{\times}_{(k,k\pm 1)}$.
Cross-order Laplacians generalize these two notions, allowing the description of diffusion processes which, as with Hodge Laplacians, can happen on simplices of any order and, as with generalized Laplacians, can \virg{jump} orders, connecting them with simplices of any other order.

\section{Statistical physics of higher-order diffusion processes}\label{section:stat_pyhs}
\paragraph{\textbf{Cross-order diffusion.}}
From now on, we will consider the high-order diffusion process on $k$-simplices through $m$-simplices on a given $\Delta$, and thus omit the $(k,m)$ notation. Such a process can be easily written as the linear, first-order ODE
\begin{equation}\label{eq:diffusion_equation}   
\Dot{x}(\tau) = -\mathbf{L}^{\times} \,x(\tau),
\end{equation}
where $x(\tau) \in \mathbb{R}^{n_k}$ is a real scalar function on the $k$-simplices.
\Cref{eq:diffusion_equation} can be solved with the time propagation operator (also called \textit{heat kernel}) at time $\tau>0$,
\begin{equation}   
\boldsymbol{\rho}(\tau) = e^{-\tau \mathbf{L}^{\times}},
\end{equation}
so that
\begin{equation}
x(\tau) = \boldsymbol{\rho}(\tau)x(0).
\end{equation}
Due to linearity, the $j$-th column $\boldsymbol{\rho}(\tau)_{,j}$ of $\boldsymbol{\rho}(\tau)$ describes the distribution of information over the $k$-simplices at time $\tau$, when the total information is concentrated in a single $k$-simplex $\sigma_j$ at time $\tau = 0$. 

It turns out that we can derive aggregate measures from the heat kernel, both for the standard Laplacian on networks ~\cite{de2016spectral} and for Hodge Laplacians on simplicial complexes \cite{baccini2022weighted}. 
These can help us probe the characteristic scales of the structure under consideration.
First,  $\boldsymbol{\rho}$ is normalized to a \emph{density  matrix} \cite{de2016spectral} 
\begin{equation}\label{eq:heat_kernel}
\Hat{\boldsymbol{\rho}}(\tau) = \frac{e^{-\tau \mathbf{L}^{\times}}}{Z(\tau)}
\end{equation}
where $Z(\tau) = \Tr(e^{-\tau \mathbf{L}^{\times}})$ is called the \emph{return probability}, describing how much of the information has remained \virg{trapped} and did not diffuse at time $\tau$ \cite{ghavasieh2022statistical}. 
With this density matrix, we can compute the von Neumann \emph{entropy} associated to the diffusion process as
\begin{equation}\label{eq:entropy}
S(\tau) = -
\Tr(\Hat{\boldsymbol{\rho}}(\tau)\log\Hat{\boldsymbol{\rho}}(\tau))
\end{equation}
and its \emph{entropic susceptibility} \cite{villegas2022laplacian}
\begin{equation}\label{eq:entropic_susceptibility}
C(\tau) = - \dfrac{\dd S}{\dd \log\tau}.
\end{equation}
The density matrix formalism and the associated von Neumann entropy allow us to describe a network's transport properties by simultaneously considering all possible diffusion trajectories. Thus, we can directly apply it to the cross-order Laplacian to investigate our hypergraph's $k$-th order properties at different scales.
This approach, in the $k = 0$ case, has been fruitfully explored in the literature to extract key information about the network's structural organization \cite{de2016spectral,ghavasieh2020,ghavasieh2023,ghavasieh2024diversity}.

In particular, the maxima and minima of the entropic susceptibility, associated with times of fast deceleration and acceleration of the diffusion process, were shown, in the case of networks, to correspond to the presence of characteristic scales~\cite{villegas2022laplacian,villegas2023laplacian}. 
Most importantly, when $C(\tau)$ is constant over a time range $I=[\tau_{\min},\tau_{\max}]$, we say that $\Delta$ exhibits informational \emph{scale-invariance}~\cite{villegas2023laplacian} in $I$. 
Explicitly writing orders $(k, m)$ again: it is known that the (network, $k=0$ and $m=1$) entropic susceptibility $C_{(0,1)}(\tau)$ shows a large plateau in the case of grid graphs, Barabási-Albert networks, and random trees, all of which are examples of self-similar structures.
Interestingly, entropic susceptibility and the associated notion of scale-invariance can be related to the concept of spectral dimension \cite{rammal1983random,ambjorn2005spectral,calcagni2014spectral}, which intuitively measures the dimensionality \virg{perceived} by a diffusion process taking place on a manifold or, in our case, a graph (see Supplementary Information Section 2).\\ 

\paragraph{\textbf{Measuring scale-invariance.}}
Following the considerations above, we want to quantify whether a higher-order network exhibits scale-invariance at order $k$ via order $m$.
To do so, we define the \emph{scale-invariance parameter} (SIP) $P_{(k,m)}(\epsilon)$, as the logarithmic lifespan of the longest connected plateau of $C_{(k,m)}(\tau)$ w.r.t. a given tolerance $\epsilon > 0$ on its ‘‘flatness''.
In detail, given a value $y>0$, we define the set
\begin{equation}
E_{(k,m)}(y;\epsilon) = \sset{\tau>0:|\log C_{(k,m)}(\tau)-y|<\epsilon}
\end{equation}
which, being the inverse image of the open interval $(y-\epsilon,y+\epsilon)$, will be given by a countable, disjoint union of open intervals $(a_i,b_i)$
\begin{equation}
    E_{(k,m)}(y;\epsilon) = \coprod_i (a_i,b_i).
\end{equation}
We then define the \emph{scale-invariance parameter} as
\begin{equation}\label{eq:scale_invariance_parameter}
P_{(k,m)}(\epsilon) = \max_{y>0} \max_i\, (\log b_i - \log a_i).
\end{equation}
If $P_{(k,m)}(\epsilon)$ is large, we can say that the hypergraph is scale-invariant at order $k$ via order $m$, while if it is close to $0$ then $C_{(k,m)}(\tau)$ is peaked, thus signaling the presence of characteristic scales.
Unless stated otherwise, in the numerical experiments we fix $\epsilon = 0.2$ and omit $\epsilon$ in the notation: $P_{(k,m)}(\epsilon) \equiv P_{(k,m)}$.

\section{Cross-order Laplacian renormalization scheme}\label{section:renormalization}
Let us assume that a hypergraph $\Delta$ has been recognized as scale-invariant with respect to $C_{(k,m)}(\tau)$ \eqref{eq:entropic_susceptibility}.
We would now like an algorithmic method to reduce it to a smaller, equivalent one that can still be recognized as scale-invariant. 
Similarly to the case of networks, we can devise a \emph{higher-order} Laplacian renormalization scheme based on the relationship between $k$-simplices through $m$-simplices. 

It consists of the following steps: 
\begin{enumerate}
    \item first, choose a diffusion order $k$ and an interaction order $m$, resulting in the cross-order Laplacian matrix $\mathbf{L}^{\times}_{(k,m)}$;
    \item choose a diffusion time $\tau^*>0$ corresponding to the scale at which to \virg{zoom out};
    \item compute a partition of the $k$-simplices from the values of $\boldsymbol{\rho}_{(k,m)}(\tau^*)$ such that simplices in the same set are strongly linked by the diffusion process at time $\tau$;
    \item coarse-grain $\Delta$ by merging its vertices according to the partition to obtain a new, smaller $\Delta'$.
\end{enumerate}
The process is then repeated, resulting in a sequence of hypergraphs with a decreasing or constant number of vertices.
We name such a sequence \emph{renormalization flow}.

Intuitively, step 3 identifies groups of $k$-simplices, which can be seen as generalizations of the spin blocks of Kadanoff's renormalization scheme \cite{kadanoff1966scaling}.
In general, however, they will not be homogeneous, but their shape will reflect the structure of $\Delta$ (see \Cref{fig:adjacency_graph}b-i).

\begin{figure*}[!htb]
    \centering
    \includegraphics[width = 0.95\linewidth]{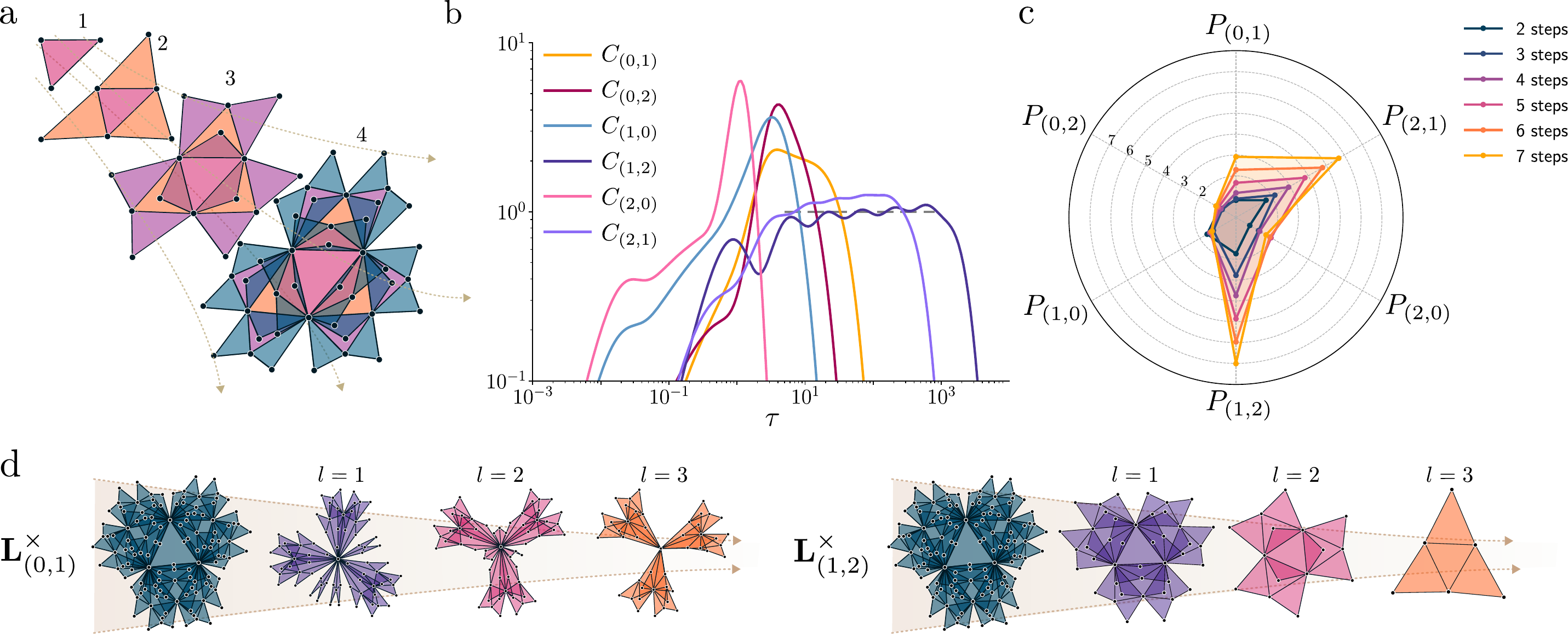}
    \caption{\textbf{Scale-invariance and renormalization in pseudofractal simplicial complexes. }{\bf a.} Graphical depiction of the first three steps of constructing the pseudofractal simplicial complex of dimension $2$. {\bf b.} Entropic susceptibility curves of all the non-zero cross-order Laplacian matrices, computed for the 2-dimensional pseudofractal simplicial complex built with six steps (1095 vertices). {\bf c.} Values of the scale-invariance parameters as the number of steps with which the pseudofractal simplicial complex is built increases. {\bf d.} On the left, the first three steps of the $\mathbf{L}^{\times}_{(0,1)}$ renormalization scheme with $\tau = 0.2$. On the right, the first three steps of the $\mathbf{L}^{\times}_{(1,2)}$ renormalization with $\tau = 2.6$. }
    \label{fig:pseudofractal_scale_invariance}
\end{figure*}

Once the partition of the $k$-simplices is obtained, we perform a coarse-graining step to aggregate the simplices belonging to the same block. 
First, we move the problem to the domain of the vertices by letting each one inherit the labels of all the $k$-simplices to which they belong (see \Cref{fig:adjacency_graph}b-ii). 
Afterward, we glue together the vertices that inherited the same label set and induce simplices from the starting $\Delta$ (see \Cref{fig:adjacency_graph}b-iii). 
A complete description of the method and a visualization of step 3 can be found in Sections 3 and 4 of the Supplementary Information, respectively.
A code implementation of the method can be found in \cite{code}.

Note that the choice of the time $\tau^*$ heavily influences the renormalization process. 
If $\tau$ is too low, no vertices will be merged, whereas if $\tau$ grows large enough, every family of connected $k$-simplices in $G_{(k,m)}$ will get the same label. 

\section{Higher-order scale-invariance}\label{section:scale_invariance}
We now show explicit examples of applications of the cross-order renormalization scheme.  
We first focus on synthetic models of simplicial complexes to confirm that, in controlled situations, the cross-order renormalization group recovers exactly the scale-invariant structure and order of the underlying system.
After this confirmation, we extract the \textit{cross-order scale signature} from some real-world datasets.

\paragraph{\textbf{Pseudofractal simplicial complexes.}}
As we mentioned above, there are situations in which the organization of a system is most evident when looking at it from a high-order point of view.
A pretty interesting example is given by the family of scale-free pseudo\-fractal simplicial complexes \cite{dorogovtsev2002pseudofractal}.
Simplicial complexes in this family are built starting with a single $k$-simplex and by iteratively attaching a $k$-simplex to each $(k-1)$-simplex already present in the complex (\Cref{fig:pseudofractal_scale_invariance}a). 

We expect the evident hierarchical nature of these objects to be visible in the entropic susceptibility curve of one or more diffusion processes defined on it.  

\Cref{fig:pseudofractal_scale_invariance}b illustrates how the different entropic susceptibilities $C_{(k,m)}$ display different non-trivial behaviors.
Most importantly, the curves associated with $\mathbf{L}^{\times}_{(1,2)}$ and $\mathbf{L}^{\times}_{(2,1)}$ show oscillating plateaux that span multiple orders of magnitude in $\tau$.
These correspond to the different, well-separated scales resulting from the iterative construction process, which can be interpreted as an approximate form of higher-order informational scale-invariance. 
Surprisingly, even if the network underlying the simplicial complex is scale-free~\cite{bianconi2016network}, the curve of $\mathbf{L}^{\times}_{(0,1)}$ (the canonical graph Laplacian) is peaked, meaning that the self-similarity of the structure is not visible at the vertex-edge level. 
\begin{figure*}[!htb]
    \centering
    \includegraphics[width = 0.9\linewidth]{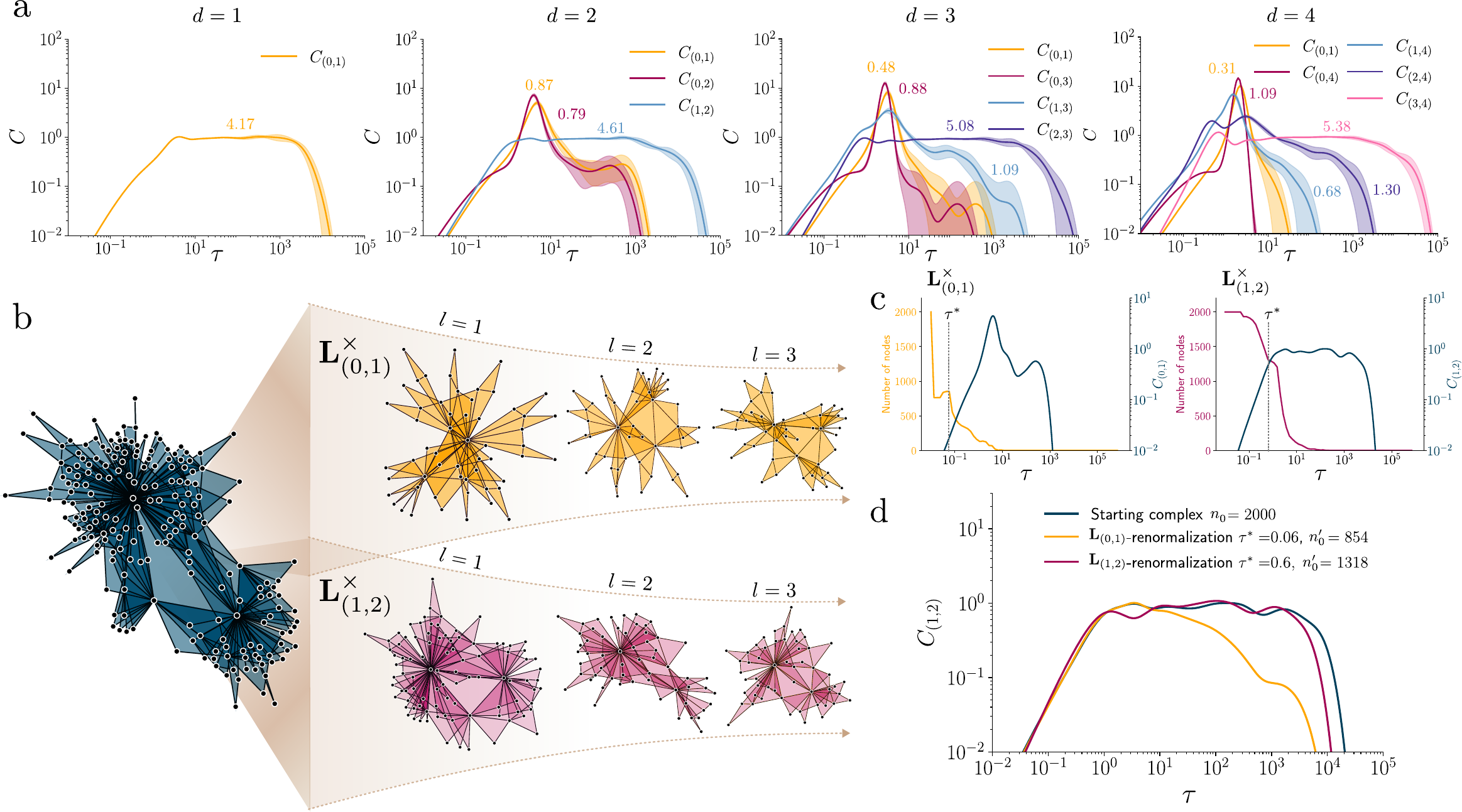}
    \caption{\textbf{Higher-order Laplacian renormalization scheme applied to heterogeneous NGF simplicial complexes} {\bf a.} Entropic susceptibility curves, together with their 95\% CIs over 10 repetitions, of the NGF simplicial complexes of flavor $s=1$, dimensions $d\in\sset{1,2,3,4}$, $\beta = 5$ and 3000 vertices. The numbers over the curve show the scale-invariance parameters. {\bf b.} A small $2$-dimensional NGF simplicial complex is renormalized using $\mathbf{L}^{\times}_{(0,1)}$ (top) and $\mathbf{L}^{\times}_{(1,2)}$ (bottom). {\bf c.} In the left panel, $C_{(0,1)}$ together with the number of vertices of the complex after one step of $\mathbf{L}^{\times}_{(0,1)}$-renormalization as a function of $\tau$. In the right panel, $C_{(1,2)}$ together with the number of vertices of the complex after one step of $\mathbf{L}^{\times}_{(1,2)}$-renormalization as a function of $\tau$.
    {\bf d.} Evolution of the entropic susceptibility $C_{(1,2)}$ over the first step of the two types of renormalization considered. The starting NGF simplicial complex has 2000 vertices and is reduced to 854 vertices by the flow of $\mathbf{L}^{\times}_{(0,1)}$ ($\tau^* = 0.06$) and to 1318 by $\mathbf{L}^{\times}_{(1,2)}$ ($\tau^* = 0.6$). }
    \label{fig:renormalizeNGF}
\end{figure*}

These results are consistently observed when we increase the number of steps with which the simplicial complex is built. 
As expected for a growing fractal structure, we find that almost all the scale-invariance parameters $P_{(k,m)}$ (\Cref{eq:scale_invariance_parameter}) for all possible cross-order Laplacians increase with the number of steps in the construction of the pseudofractal complex (\Cref{fig:pseudofractal_scale_invariance}c).
Crucially, we also find that the highest values of $P$ are consistently obtained for $P_{(1,2)}$ and $P_{(2,1)}$, which correspond to the cross-order Laplacian naturally associated with the complex's growth (addition of triangles along edges). 
Consistently, we also find that the values of $P_{(2,0)}$, $P_{(1,0)}$ and $P_{(0,2)}$ have low values which remain approximately constant, while $P_{(0,1)}$ slowly increases, indicating the presence of a growing plateau (which was invisible in \Cref{fig:pseudofractal_scale_invariance}b, where we show results for the realization with six steps). 
Finally, we observe that the observed behavior is also consistent with previous results \cite{bianconi2020spectral} on the spectrum of $\mathbf{L}^{\times}_{(0,1)}$ of pseudofractal simplicial complexes, which has power-law behavior for small eigenvalues when the number of vertices is large, i.e. a plateau in the entropic susceptibility (see Methods of \cite{villegas2022laplacian}).
However, the scale-invariant behavior is much more easily and quickly detectable when considering the entropic susceptibility of Laplacians associated with the complex's intrinsic growth process (between edges and triangles), even when the number of vertices is still small.

We confirm this observation by applying the renormalization scheme (\Cref{section:renormalization}) to a pseudofractal simplicial complex of dimension 2 and visualizing its evolution at each step.
We find that the renormalization based on $\mathbf{L}^{\times}_{(1,2)}$ preserves the structure of the simplicial complex, perfectly reversing its iterative construction process (\Cref{fig:pseudofractal_scale_invariance}d, right).
In contrast, a renormalization flow based on $\mathbf{L}^{\times}_{(0,1)}$ destroys the pseudofractal structure and rapidly collapses the central vertices, resulting in a \virg{star-shaped} simplicial complex with no apparent relation with the original one (\Cref{fig:pseudofractal_scale_invariance}d, left). 
In the Supplementary Information (Section 5), we thoroughly explore the renormalization of the pseudofractal simplicial complex under different Laplacians and values of $\tau$. 

\paragraph{\textbf{Network Geometry with flavor.}}
Up to now, we found the presence of higher-order scale-invariance only for pseudofractal simplicial complexes. 
While being a relevant consistency check, this result is not particularly striking due to the homogeneity of their structure and their evident hierarchical nature.
We show here that analogous results can be found in the case of heterogeneous simplicial complexes, similar to what was previously observed in random scale-free networks and random trees \cite{villegas2023laplacian}. To this end, a natural candidate is the family of simplicial complexes given by the \emph{Network Geometry with flavor} (NGF) model \cite{bianconi2017emergent,bianconi2016network}. 
Indeed, the NGF model in dimension $d$ can generate both hyperbolic manifolds and \emph{scale-free} growing simplicial complexes by progressively attaching $d$-simplices to $(d-1)$-simplices in a stochastic manner (see Section 6A of the Supplementary Information for details on the model). 

\begin{figure*}[!htb]
    \vspace{0.2cm}
    \centering
    \includegraphics[width=0.9\linewidth]{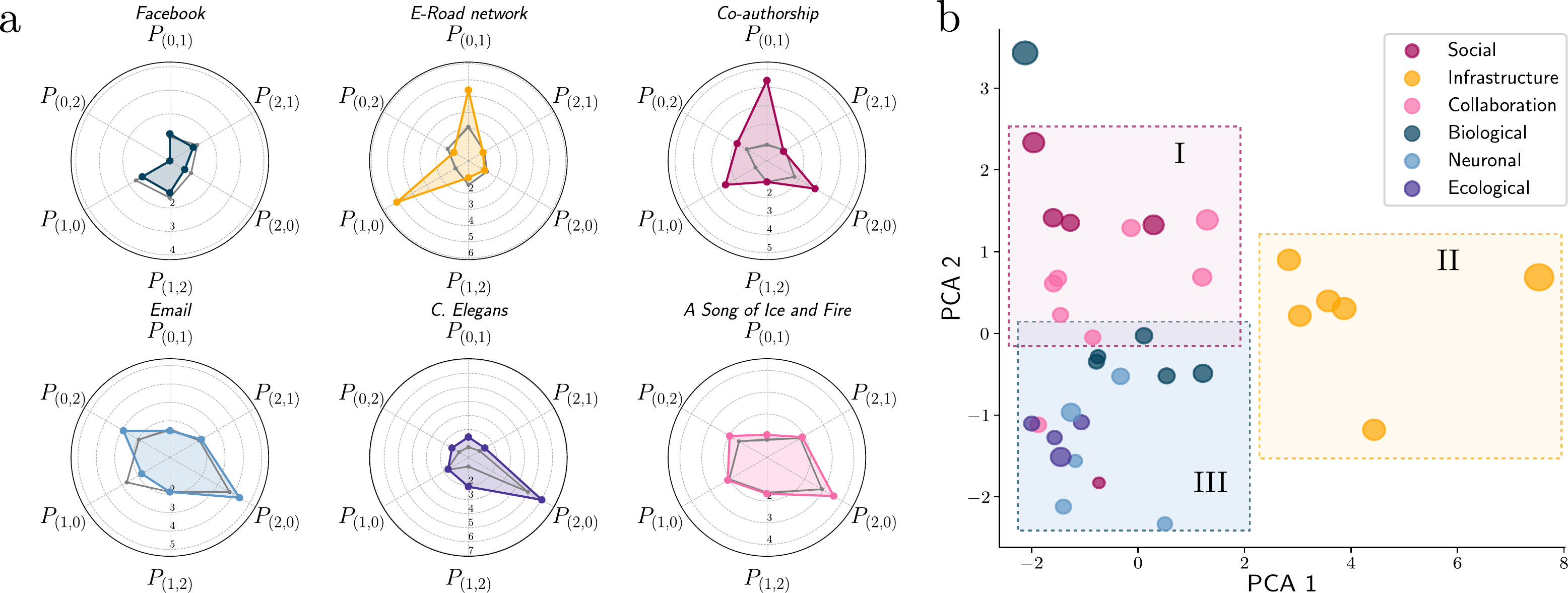}
    \caption{\textbf{Higher-order scale-invariance in real data.} {\bf a.} scale-invariance parameters $P_{(k,m)}$ computed for $k\neq m\in\sset{0,1,2}$, for six different second-order clique complexes obtained from real-world network datasets. The gray line and the associated region represent the mean and 95\% CI for the values of $P_{(k,m)}$ associated with the $(k,m)$-adjacency graphs randomized with the configuration model over 10 repetitions. {\bf b.}  The first two PCA components of a set of 34 real clique complexes of different types. The color of each point represents the associated network's type, while the size is proportional to the value of the network's highest SIP. The colored regions highlight clusters of nearby points that belong to the same class of networks: social origin (\uppercase\expandafter{\romannumeral 1\relax}), infrastructural (\uppercase\expandafter{\romannumeral 2\relax}), biological origin (\uppercase\expandafter{\romannumeral 3\relax}).}
    \label{fig:SIP_datasets}
    \vspace{-0.5cm}
\end{figure*}

This attachment process is governed by three possible \emph{flavors} $s \in \{-1, 0, 1\}$, each resulting in distinct structural properties.
Additionally, the model incorporates a parameter $\beta$ acting as an inverse temperature, influencing the randomness in the process. 

Notably, certain higher-order degrees, depending on flavor and dimension, exhibit power-law distributions, indicating a scale-free structural organization. 
Specifically, $\mathrm{deg}_{(k,d)}$ (\Cref{eq:degree}) follows a power-law when $k\leq d-3$ for $s = -1$, when $k\leq d - 2$ for $s = 0$, and for all $k\leq d-1$ for $s = 1$ (see Section 6B of the Supplementary Information for details).

Given their scale-free nature \emph{at all levels}, our focus is on NGF simplicial complexes with flavor $s = 1$. 
We examine their entropic susceptibility curves to discern scale-invariance. 
To manage computational complexity, we calculate curves only for relations predicted to be scale-free, alongside the vertex-edge $(0,1)$ relation for comparison.

The situation differs from the pseudofractal case, as shown in \Cref{fig:renormalizeNGF}a. 
Although the simplicial complex follows a similar construction process, its randomness does not allow for a clear separation of the hierarchical scales. 
Unlike the oscillating plateau observed in \Cref{fig:pseudofractal_scale_invariance}a, the curves of $C_{(d,d-1)}$ exhibit a small peak, which grows with higher dimensions, indicating the presence of a distinct microscopic scale. 
This is because the $(d-1,d)$-adjacency graph (of both $d$-dimensional NGFs and pseudofractals) is composed by $(d+1)$-cliques arranged in a tree-like structure, as $d$-simplices have $d+1$ $(d-1)$-faces.
The first peak thus corresponds to the microscale associated with these cliques (see Section 4 of the Supplementary Information for a visualization of this fact).   
Afterward, a scale-invariant regime ensues, illustrated by the near-perfect plateau of $C_{(d-1,d)}$. 
However, despite being associated with power-law degree distributions, other curves lack plateaus, showcasing non-trivial structural organization at specific scales.

In \Cref{fig:renormalizeNGF}b, we visually demonstrate that a renormalization based on a higher-order relation ($\mathbf{L}^{\times}_{(1,2)}$) better preserves the structure of a $2$-dimensional NGF simplicial complex than the standard vertex-edge Laplacian renormalization. 
Notably, \Cref{fig:renormalizeNGF}c shows that the $\mathbf{L}^{\times}_{(0,1)}$-renormalization drastically reduces the number of vertices, collapsing them into a single super-vertex at the first peak of $C_{(0,1)}$. 
Conversely, $\mathbf{L}^{\times}_{(1,2)}$-renormalization compresses the simplicial complex more gradually, revealing a clear transition point just before the first peak. 
This distinction is further emphasized by tracking the evolution of the entropic susceptibility after one step of the renormalization flow in \Cref{fig:renormalizeNGF}d, where the plateau of $C_{(1,2)}$ is preserved by $\mathbf{L}^{\times}_{(1,2)}$ but destroyed by $\mathbf{L}^{\times}_{(0,1)}$.

\paragraph{\textbf{Higher-order scale-invariance in real data.}}
The framework we established can be fruitfully employed to scrutinize the structure of real-world networks.
We can leverage scale-invariance parameters to provide effective higher-order signatures describing their hierarchical nature in a multifaceted way. 
We take 6 different real network datasets (see Section 7 of the Supplementary Information for details) and consider their associated \emph{clique complexes}, i.e. build simplicial complexes by filling their cliques up to the $2$nd order.
For each, we compute all the $C_{(k,m)}$ and the associated $P_{(k,m)}$ and build a null model by doing the same on their adjacency graphs randomized with a configuration model.
As shown in \Cref{fig:SIP_datasets}a, we obtain an effective signature of the network encoding the amount of scale-invariance in each of its higher-order relations.
We find a zoo of different behaviors, like the \emph{E-road} network \cite{vsubelj2011robust} and the network science \emph{Co-authorship} network \cite{newman2006finding}, which are strongly scale-invariant in the standard sense, i.e. in the relation between vertices and edges (high values of $P_{(0,1)}$ and $P_{(1,0)}$).
Others, like the \emph{Facebook} network \cite{leskovec2012learning} show negligible values of scale-invariance in every order.
Interestingly, despite the inherent pairwise nature of the networks, we find a marked presence of high-order scale-invariance in some of them, like the University Rovira i Virgili \emph{Email} network \cite{guimera2003self}, the \emph{C.Elegans} metabolic network \cite{duch2005community,jeong2000large,overbeek2000wit} and the fictional \emph{A Song of Ice and Fire} social network \cite{asoiaf}. 
This higher-order mark, notice, is obtained without any assumptions on the actual higher-order interactions present, if any, in the networks but is based only on nodes, edges, and the cliques they form.
This observation validates our claim that some aspects of the hierarchical scale-invariant structure of a higher-order network may be hidden from the node-centric point of view but can be evident when looking at how simplices (in this case, cliques) are related to one another.

Finally, we use scale-invariance parameters as coordinates to embed a larger set of real-world second-order clique complexes in $\mathbb{R}^6$ (for details, see Supplementary Information Section 7C).
The first two principal components of the resulting point cloud, we see in \Cref{fig:SIP_datasets}b, show high heterogeneity in the amount of higher-order scale-invariance present in each dataset.
Surprisingly, despite the coarseness of this measure, we see that networks of similar nature tend to be closer to each other, as shown by the colored regions, highlighting the presence of specific higher-order signatures associated with each type of data.

\section{Discussion}
\label{sec:discussion}
By directly extending the Laplacian renormalization group framework to higher-order interactions, we developed a method to investigate the structure of higher-order networks via a robust renormalization procedure based on the connectivity structure of higher-order relations, as encoded in the proposed cross-order Laplacian.  
We showed that the renormalization scheme can revert exactly the construction of simplicial structures that are self-similar by construction (the pseudofractal) and detect the correct order of the dominating growth mechanism in scale-invariant complexes with induced heterogeneous lower order structures (the NGF model). 
Armed with these results on controlled synthetic systems, we leveraged the entropic susceptibilities obtained in our scheme to build \textit{scale-invariance profiles} for a set of real-world systems, revealing both different unexpected kinds of characteristic scales and scale-invariance at various orders and commonality of such profiles across systems belonging to the same domain. 

From a technical point of view, the cross-order Laplacians bear both similarities and differences with previously defined higher-order Laplacians, notably the Hodge (or combinatorial) Laplacian~\cite{muhammad2006control} $\boldsymbol{\mathcal{L}}_k$, and the multi-order Laplacian $\mathbf{L}^{(\mathrm{mul})}$ \cite{lucas2020multiorder}.  
In fact, both the $k$-order Hodge Laplacian and cross-order Laplacians $\mathbf{L}^{\times}_{(k,m)}$ (for any $m$) are defined on the simplices of order $k$, and thus are $n_k \cross n_k$ matrices, where $n_k$ is the number of $k$-simplices.
However, the Hodge Laplacian is limited to adjacencies defined by boundary and coboundary relations, and thus, with $m=k\pm1$, while the cross-order can capture arbitrary adjacencies through any $m$-simplices.
Conversely, the multi-order Laplacian $\mathbf{L}^{(\mathrm{mul})}$ is defined as the weighted sum of Laplacians defined on nodes and adjacent via simplices of any $m$ order. 
It can be rewritten in our notation as $\mathbf{L}^{(\mathrm{mul})} = \sum_{m=1}^M \omega_m \mathbf{L}^{\times}_{(0,m)}$, where the $\omega_m$ are an arbitrary weighting scheme and $M$ the maximal order considered. 
Similarly, we note that, while in this work we focused on renormalization based on specific pairs $(k, m)$, important future work should understand the effect of extending the cross-order renormalization scheme to a \textit{multicross}-order scheme, which could be achieved by considering combinations of simplices of dimension higher or lower than a certain threshold on the interaction order (e.g., $\mathbf{L}^{(\mathrm{mul}\times)}_k = \sum_{m \in \{\mathbf{m}\} } \omega_{m} \mathbf{L}^{\times}_{(k,m)}$ over a set $\{\mathbf{m}\}$). 
Also, while this work focused on simplicial complexes for convenience and clearness of exposition, any combinatorial structure \cite{hajijtopological} built with a set of ranked sets is amenable to our scheme.

Finally, our results provide a new lens to address questions on the origin of different higher-order invariant structures in various domains~\cite{thibeault2024low,lyu2024learning}, and on their effects on dynamical processes taking place on them~\cite{ferraz2023multistability,ferraz2023multistability,simplicial_contagion}, as well as to the limits they might pose on the predictability and reconstruction of complex systems \cite{murphy2022duality}.

\section{Acknowledgements}
M.N. acknowledges the project PNRR-NGEU, which has received funding from the MUR – DM 352/2022. T.G. thankfully acknowledges financial support by the European Union - NextGenerationEU - National Recovery and Resilience Plan (Piano Nazionale di Ripresa e Resilienza, PNRR), project `SoBigData.it - Strengthening the Italian RI for Social Mining and Big Data Analytics' - Grant IR0000013 (n. 3264, 28/12/2021). 
We also thank P. Villegas and A. Gabrielli for extremely valuable suggestions on preliminary
versions of the manuscript.

\bibliography{aipsamp3}



\section*{Supplementary material}

# I. RELATION WITH THE HODGE LAPLACIAN MATRICES

The combinatorial Hodge Laplacian matrix [1], first introduced by Eckmann [2], is the combinatorial analogue of the Hodge Laplacian acting on $p$-forms of differential geometry. It is a $n_p \times n_p$ matrix which describes a particular relation between $p$-simplices which depends on $(p-1)$ and $(p+1)$-simplices. To build it, we first need to assign an *orientation* to each simplex $\sigma = \{v_0, \ldots, v_k\} \in \Delta$, i.e. an ordering of its vertices $\sigma = [v_0, \ldots, v_k]$. When the simplices are oriented, it is possible to define a notion of coherence and incoherence of orientation between $(p+1)$ and $(p-1)$-adjacent $p$-simplices (see [3] for details). We write $\sigma \sim \eta$ when $\sigma, \eta$ are coherently oriented and $\sigma \nsim \eta$ when they are incoherently oriented. The Hodge-Laplacian is then defined as

$$(\boldsymbol{\mathcal{L}}_p)_{ij} = \begin{cases} \mathbf{k}_{p,p+1}(\sigma_i) + p + 1 & \text{if } i = j \\ 1 & \text{if } i \neq j, \ a_{(p,p-1)}(\sigma_i, \sigma_j) \neq 0, \ a_{(p,p+1)}(\sigma_i, \sigma_j) = 0 \ \text{and } \sigma_i \sim \sigma_j \ , \\ -1 & \text{if } i \neq j, \ a_{(p,p-1)}(\sigma_i, \sigma_j) \neq 0, \ a_{(p,p+1)}(\sigma_i, \sigma_j) = 0 \ \text{and } \sigma_i \nsim \sigma_j \end{cases} \tag{1}$$

where $\mathbf{k}_{p,p+1}(\sigma_i)$ is the number of $(p+1)$-simplices which contain $\sigma_i$ and $a_{(p,p\pm1)}$ is the adjacency number as defined in the main text. The pattern of non-zero elements of Equation (1) makes it clear how $\boldsymbol{\mathcal{L}}_p$ describes relations among $p$-simplices which are adjacent from below but are *not* adjacent from above. When two $p$-simplices $\sigma, \eta$ satisfy this condition, they are said to be *parallel neighbors* [4] and we write $\sigma \parallel \eta$.

Moreover, there are other notable differences between this matrix and the standard graph Laplacian. First, the oriented nature of the simplices is such that the extra-diagonal elements can be both $-1$ and $+1$, instead of being all $-1$. Second, the rows in general do not sum to zero. This last property comes from the fact that the vector $\mathbb{1} = (1, \ldots, 1)^\top$ does not belong in general to $\ker \boldsymbol{\mathcal{L}}_p$, which instead contains the so-called *harmonic vectors* corresponding to the $p$-th homology classes of $\Delta$.

These two properties make it so that diffusion through the Hodge-Laplacian is not a standard "intuitive" diffusion process. The fact that $\mathbb{1} \notin \ker \boldsymbol{\mathcal{L}}_p$, for instance, means that the total information present in the $p$-simplices is *not* conserved in time, as, under the dynamics $\dot{x}(\tau) = -\boldsymbol{\mathcal{L}}_p x(\tau)$

$$\frac{\mathrm{d}}{\mathrm{d}\tau} \sum_i x_i(\tau) = \frac{\mathrm{d}}{\mathrm{d}\tau} \mathbb{1}^\top x(\tau) = \mathbb{1}^\top \dot{x}(\tau) = \mathbb{1}^\top \boldsymbol{\mathcal{L}}_p x(\tau) \neq 0.$$

The Hodge Laplacian and the graph-like cross-order Laplacian defined in the main text, however, can be related in the following way. In Ref. [4], Forman introduces a combinatorial version of the well-known Weitzenböck identity of Riemannian geometry, which states that the Hodge Laplacian can be decomposed as the sum of the Bochner (or rough) Laplacian and a term depending only on curvature. In our discrete setting this amounts to

$$\boldsymbol{\mathcal{L}}_p = \boldsymbol{\mathcal{L}}_p^B + \mathbf{F}_p, \tag{2}$$

where $\boldsymbol{\mathcal{L}}_p^B$ is the $p$-th *Bochner Laplacian*, a positive semidefinite matrix defined as

$$(\boldsymbol{\mathcal{L}}_p^B)_{ij} = \begin{cases} \sum_{l \neq i} |\boldsymbol{\mathcal{L}}_p|_{il} & \text{if } i = j \\ (\boldsymbol{\mathcal{L}}_p)_{ij} & \text{if } i \neq j \end{cases}, \tag{3}$$

and $\mathbf{F}_p$, named *Forman curvature*, is a diagonal matrix containing the combinatorial curvatures of the $p$-simplices. This particular decomposition is such that $\mathbf{F}_p$ can be thought of as a "correction" to the fact that the absolute values of the row elements of $\boldsymbol{\mathcal{L}}_p$ do not sum to 0. We now have that the diagonal elements of the Bochner Laplacian correspond to the number of non-zero elements in each row, i.e. the number of parallel neighbors

$$(\boldsymbol{\mathcal{L}}_p^B)_{ii} = | \{ \sigma_j \in \Delta_p \mid \sigma_i \parallel \sigma_j \} | \tag{4}$$

which, given that in a simplicial complex simplices adjacent from above are also adjacent from below, can in turn be related to the higher-order degrees defined in the main text:

$$(\boldsymbol{\mathcal{L}}_p^B)_{ii} = \deg_{(p,p-1)}(\sigma_i) - \deg_{(p,p+1)}(\sigma_i). \tag{5}$$

This tells us that we can write the $p$-th Bochner Laplacian as the difference of two cross-order Laplacians multiplied element-wise with a "residual" matrix $\mathbf{R}_p$ which includes the effects of orientations:

$$\boldsymbol{\mathcal{L}}_p^B = (\mathbf{L}_{(p,p-1)}^\times - \mathbf{L}_{(p,p+1)}^\times) \odot \mathbf{R} = \mathbf{L}_p^\parallel \odot \mathbf{R}_p, \tag{6}$$

where $(\mathbf{R}_p)_{ij} = -1$ if $\sigma_i \parallel \sigma_j$ and $\sigma_i \sim \sigma_j$ and 1 otherwise. Moreover, it is possible to interpret $\mathbf{L}^\times_{(p,p-1)} - \mathbf{L}^\times_{(p,p+1)}$ as the Laplacian of a new adjacency graph, which we name *parallel adjacency graph*, where nodes represent $p$-simplices and edges connect parallel neighbors, hence the notation $\mathbf{L}^\parallel_p$.

Finally, putting together Equation (2) with Equation (6), we find

$$\mathcal{L}_p = \mathbf{L}^\parallel_p \odot \mathbf{R}_p + \mathbf{F}_p,\tag{7}$$

which suggests that we may interpret diffusion with the Hodge Laplacian, neglecting the contribution of orientation, as a reaction-diffusion process where diffusion takes place among parallel $p$-simplices and reaction, responsible for the production and destruction of information, is given by the Forman curvature.

## II. SCALE-INVARIANCE AND SPECTRAL DIMENSION

To better understand the meaning of the definition of scale-invariance employed in this work, we can look from two different angles. First, we notice,

$$C(\tau) = C^* \iff \frac{\mathrm{d}C(\tau)}{\mathrm{d}\log\tau} = 0 \iff \frac{\mathrm{d}^2 S(\tau)}{\mathrm{d}(\log\tau)^2} = 0 \quad \forall \tau \in I$$

which means that scale-invariance is associated to a range of times where the rate of change of the entropy (in logarithmic scale) is constant, i.e. $S(\tau)$ does not accelerate nor decelerate. From another point of view, it is interesting to look at the relation between the entropic susceptibility and the *spectral dimension* of the adjacency graphs of the simplicial complex. The spectral dimension [5–7] intuitively measures the dimensionality "perceived" by a diffusion process taking place on a manifold or, in our case, a graph. We define the spectral dimension $D_s(\tau)$ as the derivative w.r.t. the logarithmic diffusion time of the logarithm of the return probability $Z(\tau)$

$$D_s(\tau) = -2\frac{\mathrm{d}\log Z(\tau)}{\mathrm{d}\log\tau}.\tag{8}$$

Here $\tau$ has to be thought of as a *scale* instead of a time parameter, meaning that $D_s(\tau)$ measures dimensionality at scale $\tau$. On $d$-dimensional flat manifolds $D_s(\tau) = d$ for all $\tau$, while on $d$-dimensional (periodic) lattices $D_s(\tau)$ shows a large plateau whose value corresponds exactly to $d$ [7]. In general, when $Z(\tau) \propto \tau^{-2d}$ in a scale interval $I = [\tau_{\min}, \tau_{\max}]$, $D_s(\tau)$ has a plateau equal to $d$. This implies that, at these specific temporal resolutions, the spectral dimension observed through the diffusion process remains consistent. Such behavior suggests the existence of an inherent dimensionality within the space at these particular scales.

One can see that there is a strong relation between the entropic susceptibility

$$C(\tau) = -\frac{\mathrm{d}S}{\mathrm{d}\log\tau}$$

and the spectral dimension, i.e.

$$C(\tau) = -\frac{1}{2}\frac{\mathrm{d}D_s(\tau)}{\mathrm{d}\log\tau} + \frac{1}{2}D_s(\tau).\tag{9}$$

From Equation (9), we thus find that scale-invariance is equivalent to

$$C(\tau) = C^* \iff D_s(\tau) = a\tau + 2C^*\tag{10}$$

i.e. the spectral dimension varies linearly w.r.t. the scale. This result tells us that informational scale-invariance corresponds not only to spaces with a well-defined intrinsic dimensionality ($a = 0$), but also the case in which the dimension varies linearly with the scale ($a \neq 0$). This last situation, as we can see by taking the definition of spectral dimension (8) and integrating Equation (10), corresponds to

$$Z(\tau) \propto e^{-\frac{a}{2}\tau}\tau^{-C^*},\tag{11}$$

i.e. changing the scale results in the space becoming increasingly "larger" so that the return probability decreases exponentially fast.

### III. DETAILS ON THE RENORMALIZATION METHOD

Here we explain in detail our higher-order renormalization algorithm when applied to a general hypergraph $\Delta$. The main steps can be outlined in the following way:

1. first, choose a diffusion order $k$ and an interaction order $m$, resulting in the cross-order Laplacian matrix $\mathbf{L}_{(k,m)}^{\times}$;

2. choose a diffusion time $\tau^* > 0$ corresponding to the scale at which to "zoom out";

3. compute a partition of the $k$-simplices from the values of $\boldsymbol{\rho}_{(k,m)}(\tau^*)$ such that simplices in the same set are strongly linked by the diffusion process at time $\tau$;

4. coarse-grain $\Delta$ by merging its vertices according to the partition, in order to obtain a new, smaller hypergraph $\Delta'$.

In detail, staying faithful to the real-space Laplacian renormalization scheme, we say that $k$-simplices $\sigma$ and $\eta$ are to be merged if the information which has flowed from $\sigma$ to $\eta$ is greater than the information which has stayed in $\sigma$ or $\eta$. In practice, we build an auxiliary matrix $\boldsymbol{\zeta}$ with

$$\boldsymbol{\zeta}_{ij} = \begin{cases} 1 & \text{if } \boldsymbol{\rho}_{ij} \geq \min(\boldsymbol{\rho}_{ii}, \boldsymbol{\rho}_{jj}) \\ 0 & \text{otherwise} \end{cases} \tag{12}$$

which can be thought as the adjacency matrix of a *metagraph* $\tilde{G}$ where the nodes represent the $k$-simplices and the edges connect those to be identified w.r.t. the diffusion process. A partition of the $k$-simplices is then obtained simply by taking the connected components of $\tilde{G}$.

As prescribed by step 4, we need a coarse grain of $\Delta$ which is informed by the partition obtained in step 3.

When the diffusion takes place between the vertices of a network, the process is quite simple [8]. It is enough to glue together vertices belonging to the same set into *super-vertices*, and place a *super-edge* $\{A, B\}$ between two super-vertices $A$ and $B$ if there are two vertices $a \in A$ and $b \in B$ in the original network such that $\{a, b\} \in \Delta_1$. This is readily generalized to the case of a higher-order network $\Delta$ renormalized with a Laplacian of the form $(0, m)$. In fact, we can induce an $l$-th order *super-simplex* $\{A_0, \ldots, A_l\}$ between super-vertices $A_0, \ldots, A_l$ if there are vertices $a_0 \in A_0, \ldots, a_l \in A_l$ such that $\{a_0, \ldots, a_l\} \in \Delta$.

If we want to renormalize the higher-order network w.r.t. a process taking place on general $k$-simplices, more care is needed, as it is not clear how to naturally collapse simplices belonging to the same set in the partition. It is important to highlight that we cannot just coarse grain the adjacency graph w.r.t. the partition, as that will result in a smaller graph where most of the relational information contained in the hypergraph has been lost. To approach the problem, we propose the following method which, taking inspiration from Ref. [9], aims to reduce the problem to a node coarse-graining one.

Let us suppose that each $k$-simplex $\eta$ has been assigned a label $\Lambda(\eta)$ specifying which set of the partition it belongs to. First, each $k$-simplex passes its label down to its vertices, or, from the other point of view, each vertex inherits the labels from all the $k$-simplices to which it belongs. Thus, we associate to each vertex $v$ a set, called *signature* $\Sigma(v)$,

$$\Sigma(v) = \{\Lambda(\eta) \mid \eta \in \Delta_k, \ v \in \eta\}, \tag{13}$$

which contains all the labels inherited. The coarse graining is then performed simply by identifying vertices possessing the same signature and then inducing higher-order simplices as explained above. If $\Sigma(v) = \emptyset$, meaning that $i$ is not contained in any $k$-simplex, then we give it its own signature, so that it will be left unchanged by the transformation.

In other terms, we think of the signature as a labelling of the vertices of the hypergraph, i.e. a surjective map $f : \Delta_0 \to S$ from the set of vertices to the set of signatures $S$. As it is explained in Ref. [3] for simplicial complexes, starting from this labelling, we can build another hypergraph $\Delta'$ whose vertices are the signatures, and whose simplices consist of the maps of the simplices of $\Delta$ through $f$, i.e.

$$\Delta' = \{\{f(v_0), \ldots, f(v_k)\} \mid \{v_0, \ldots, v_k\} \in \Delta_k, \ k \in \mathbb{N}\}. \tag{14}$$

## IV. SIMPLICIAL PARTITION AS A FUNCTION OF TIME

In Figure S1, we visually display the behavior of steps 2,3 of the renormalization method described in the Section III, applied on a 3-dimensional NGF simplicial complex $\Delta$ of 220 vertices with $\beta = 0.1$. In the top part of the figure, we show the entropic susceptibility $C_{(2,3)}$ associated to the diffusion process where information is situated on triangles (2-simplices) and flows through tetrahedra (3-simplices). As discussed in the main text, this simplicial complex is $(2,3)$-scale-invariant, as we see in the large plateau which spans multiple orders of magnitude of $\tau$. We pick 4 different diffusion times $\tau^* \in \{0.6, 5, 100, 6000\}$, where the first one corresponds to the first peak in $C_{(2,3)}$ and the last one falls beyond the plateau. In the second row, we see the simplicial complex $\Delta$ with its triangles (2-simplices) colored according to the partition obtained at different times $\tau^*$. In the third row, we see the associated $(2,3)$-adjacency graph with its nodes again colored according to the partition at each time.

As we can see, the $(2,3)$-adjacency graph of a 3-dimensional NGF is composed of cliques of 4 nodes (corresponding to the 4 faces in a single tetrahedron) connected together through single nodes in a tree-like fashion. The first peak in the entropic susceptibility indeed corresponds to the micro-scale associated to the integration of information in these fundamental cliques. In fact, when $\tau^* = 0.6$, the partition of the nodes of $G_{(2,3)}$ tends to assign the same label to nodes in the same clique. As the diffusion time increases, the number of sets in the partition decreases and each set identifies bigger and bigger branches. When $\tau^*$ reaches 6000, we see in the rightmost panel, we reach full information integration and all nodes (2-simplices) belong to a single set.

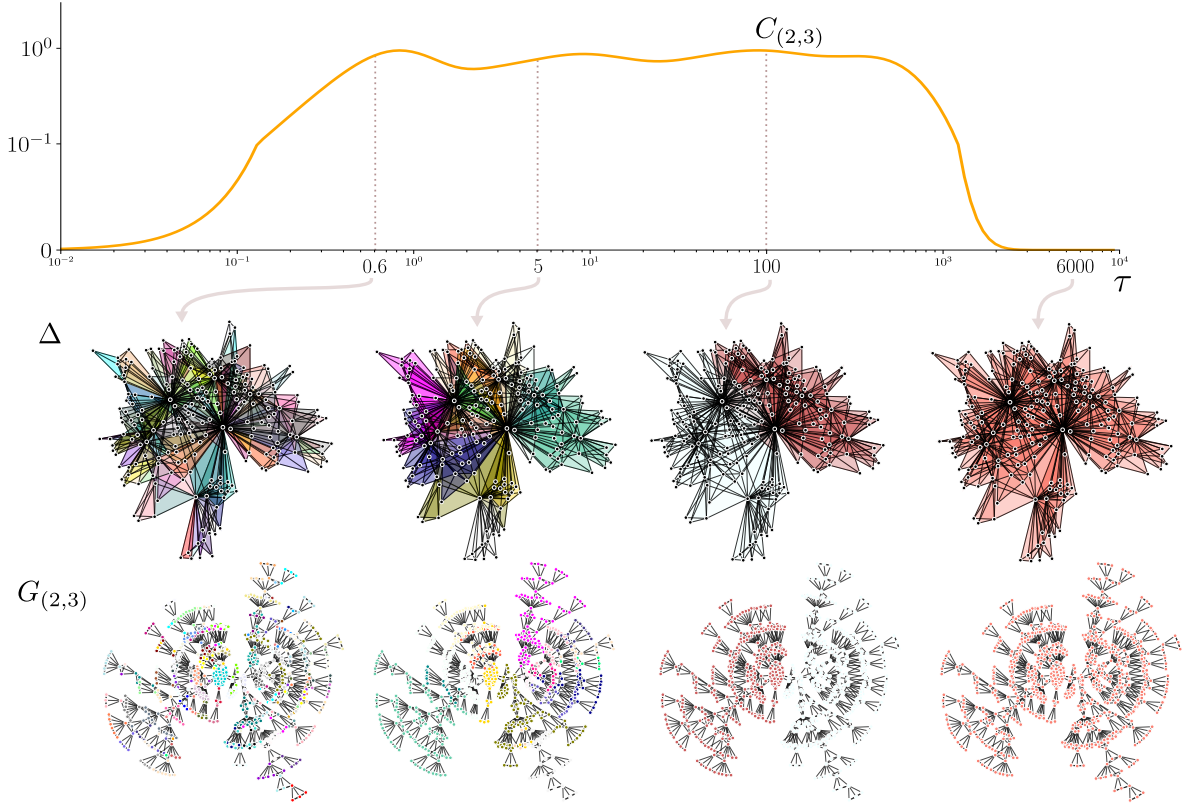

FIG. S1. Simplices partition as a function of diffusion time.

## V. RENORMALIZATION OF THE PSEUDOFRACTAL SIMPLICIAL COMPLEX

As discussed in the main text, in some simplicial complexes the constraints imposed by a higher-order coarse graining scheme are essential to have a meaningful reduction of their structure. In Figures S2 and S3, we see the 2-dimensional pseudofractal simplicial complexes renormalized with all cross-order Laplacians $\mathbf{L}_{(k,m)}^{\times}$ $(k, m \in \{0, 1, 2\}, \ k \neq m)$, each with 10 different diffusion times $\tau$. The renormalization with $\mathbf{L}_{(1,2)}^{\times}$ is the only one capable of disentangling the characteristic scales and transform the pseudofractal in a pseudofractal of the same type ($\tau = 2.22$ and $\tau \geq 5.56$ in the figure).

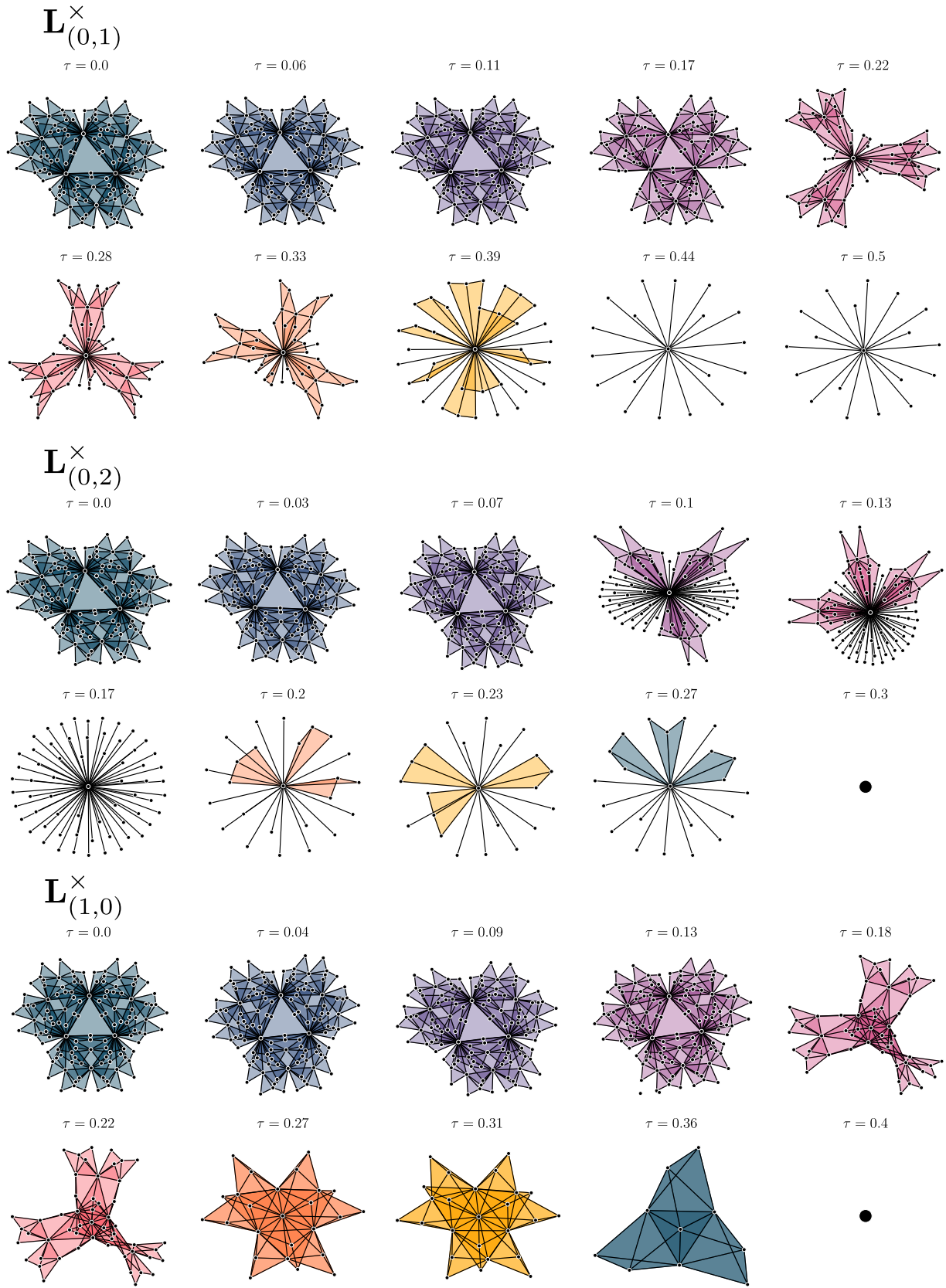

FIG. S2. Renormalization of the 2-dimensional pseudofractal simplicial complex using cross-order Laplacians $\mathbf{L}^{\times}_{(0,1)}$, $\mathbf{L}^{\times}_{(0,2)}$ and $\mathbf{L}^{\times}_{(1,0)}$.

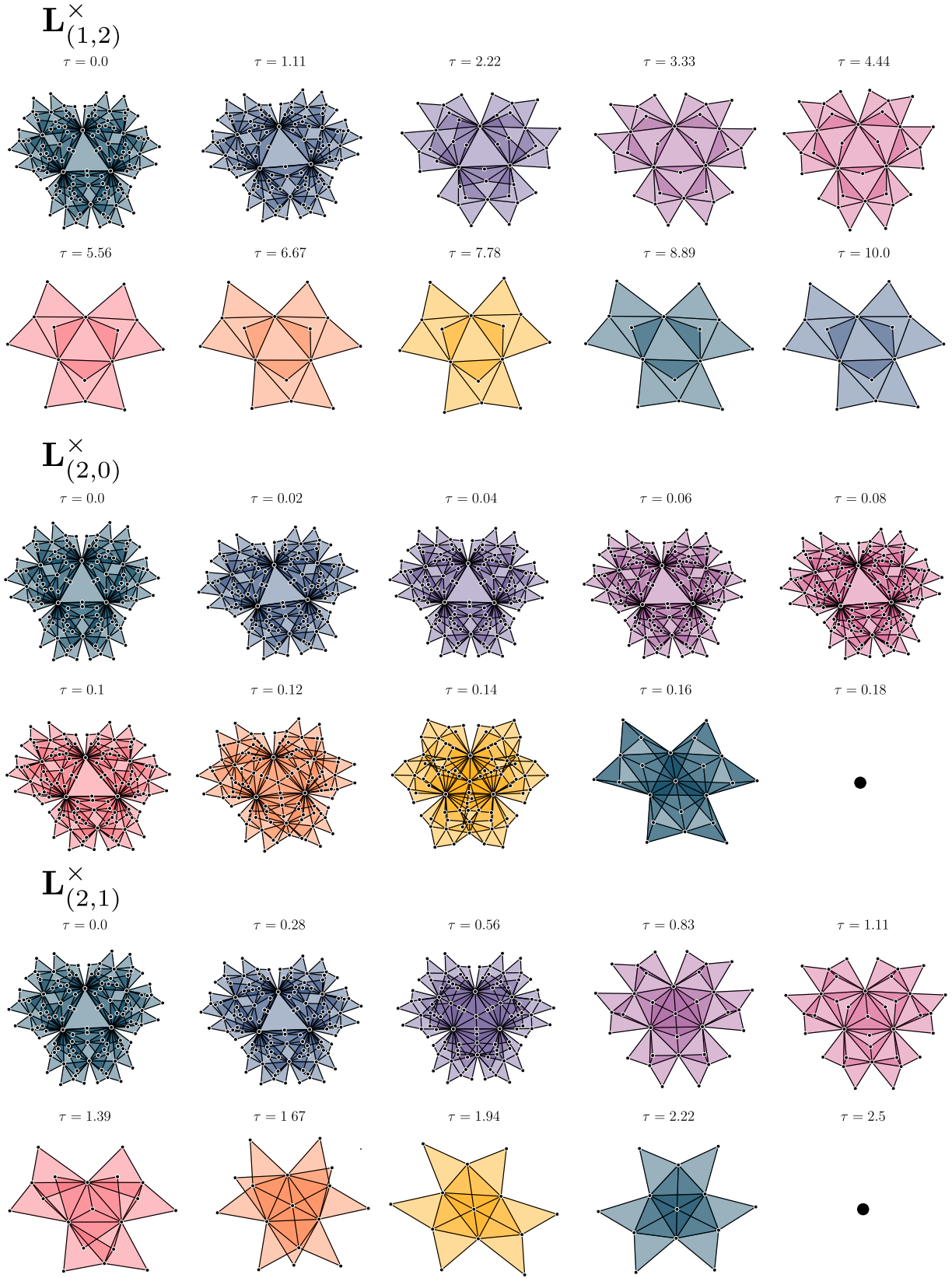

FIG. S3. Renormalization of the 2-dimensional pseudofractal simplicial complex using cross-order Laplacians $\mathbf{L}^{\times}_{(1,2)}$, $\mathbf{L}^{\times}_{(2,0)}$ and $\mathbf{L}^{\times}_{(2,1)}$.

## VI.   RENORMALIZATION OF NETWORK GEOMETRY WITH FLAVOR

### A.   Model

The NGF model [10] in dimension $d$ is able to generate both hyperbolic manifolds and *scale-free* growing simplicial complexes, by progressively attaching $d$-simplices to $(d-1)$-simplices in a stochastic manner biased by the flavor $s$. The three possible ways in which the growth process is realized, named *flavors*, are the following:

- $s = -1$, which allows for at most 2 $d$-simplices attached to any $(d-1)$-simplex, resulting in $d$-dimensional simplicial manifolds;

- $s = 1$, which glues $d$-simplices to $(d-1)$-simplices by preferential attachment;

- $s = 0$, which presents intermediate properties between the two.

Let us briefly describe the growth process of the model.

At time $t = 1$ the NGF simplicial complex $\Delta^{(1)}$ is made by a single $d$-simplex. At each time step $t > 1$ a new $d$-simplex is created and attached to one of the $(d-1)$-simplices present in $\Delta^{(t)}$, chosen with probability

$$\Pi_\sigma^{[s]} = \frac{1}{Z^{[s]}(t)} e^{-\beta \epsilon_\sigma} (1 + s n_\sigma),\tag{15}$$

for each $(d-1)$-simplex $\sigma$, where

- $\beta$ is the inverse temperature, which controls the amount of randomness in the process;

- $\epsilon_\sigma$ is the *energy* of the simplex $\sigma$, defined as the sum of the energies of its vertices, which in turn are sampled from a distribution $g(\epsilon)$;

- $n_\sigma$ is the number of $d$-simplices which contain $\sigma$ minus one;

- $Z^{[s]}(t)$ is the normalization constant.

From this it is easy to see that, when $s = -1$, all the $(d-1)$-simplices which are contained in exactly 2 $d$-simplices will have probability 0 to have another simplex attached to them. This results in $\Delta^{(t)}$ being a simplicial manifold (with boundary) for every $t \geq 1$.

### B.   Higher-order degrees

In Ref. [10], it is proven that in a $d$-dimensional NGF simplicial complex with flavor $s = 1$, the generalized higher-order degrees associated to $m$-simplices are power-law distributed (when $m \leq d-1$). In that work, however, the definition of higher-order degree differs from the one we employ i.e.

$$\deg_{(k,m)}(\sigma) = \sum_{\eta \in \Delta_k} a_{(k,m)}(\sigma, \eta).\tag{16}$$

In particular, the higher-order degree $\mathbf{k}_{m,d}(\sigma)$ of the $m$-simplex $\sigma \in$ is defined as the number of $d$-simplices which contain $\sigma$, i.e.

$$\mathbf{k}_{m,d}(\sigma) = |\{\eta \in \Delta_d : \ \sigma \subset \eta\}|.\tag{17}$$

In this section, we prove that the degrees defined in these two ways are proportional, with a proportionality constant dependent only on $m$ and $d$. This means that when $\mathbf{k}_{m,d}$ is power-law distributed, then the same holds for $\deg_{(m,d)}$.

**Proposition 1** *If $m < d$ and $\sigma \in \Delta_m$, then*

$$\deg_{(m,d)}(\sigma) = \left[ \binom{d+1}{m+1} - 1 \right] \mathbf{k}_{m,d}(\sigma).\tag{18}$$

**Proof 1** *It holds that*

$$\deg_{(m,d)}(\sigma) = \sum_{\tau \in \Delta_m} a_{(m,d)}(\tau, \sigma) = \sum_{\tau \in \Delta_m, \tau \neq \sigma} |\{\eta \in \Delta_d : \sigma \cup \tau \subseteq \eta\}| = \sum_{\tau \in \Delta_m, \tau \neq \sigma} \sum_{\eta \in \Delta_d} b(\sigma; \eta) b(\tau; \eta)$$

*where we define $b(\sigma; \eta) = 1$ if $\sigma \subset \eta$ and 0 otherwise. It follows that*

$$\deg_{(m,d)}(\sigma) = \sum_{\tau \in \Delta_m, \tau \neq \sigma} \sum_{\eta \in \Delta_d} b(\sigma; \eta) b(\tau; \eta) = \sum_{\eta \in \Delta_d} b(\sigma; \eta) \sum_{\tau \in \Delta_m, \tau \neq \sigma} b(\tau; \eta) = \sum_{\eta \in \Delta_d, \sigma \subset \eta} |\{\tau \in \Delta_m : \tau \neq \sigma, \ \tau \subset \eta\}|.$$

*Given that $\Delta$ is a simplicial complex, every face of a $d$-simplex is a simplex in the simplicial complex. In particular, a $d$-simplex has exactly $\binom{d+1}{m+1}$ faces of order $m$ and thus*

$$|\{\tau \in \Delta_m : \tau \neq \sigma, \ \tau \subset \eta\}| = \binom{d+1}{m+1} - 1. \tag{19}$$

*We can thus conclude the proof by noticing that*

$$\deg_{(m,d)}(\sigma) = \left[\binom{d+1}{m+1} - 1\right] \sum_{\eta \in \Delta_d, \sigma \subset \eta} 1 = \left[\binom{d+1}{m+1} - 1\right] \mathbf{k}_{m,d}(\sigma).$$

## VII. RENORMALIZATION OF REAL SIMPLICIAL COMPLEXES

### A. Entropic susceptibilities after renormalization

As discussed in the main text, we first considered 6 second-order clique complexes obtained from real world dataset taken from the KONECT project [11]. We renormalize each one of them using the cross-order Laplacian associated to the highest scale-invariance parameter together with $\mathbf{L}_{(0,1)}^{\times}$ as a reference, choosing the smallest time $\tau^{*}$ such that the number of nodes is reduced by at least 40%. In Figure S4 the entropic susceptibility associated to the highest SIP is shown, before and after renormalization. We highlight that the metabolic network of the C. Elegans is the only scale-invariant one in which we can observe that the renormalization with $\mathbf{L}_{(0,1)}^{\times}$ better preserves the plateau than $\mathbf{L}_{(2,0)}^{\times}$, despite the SIP being lower.

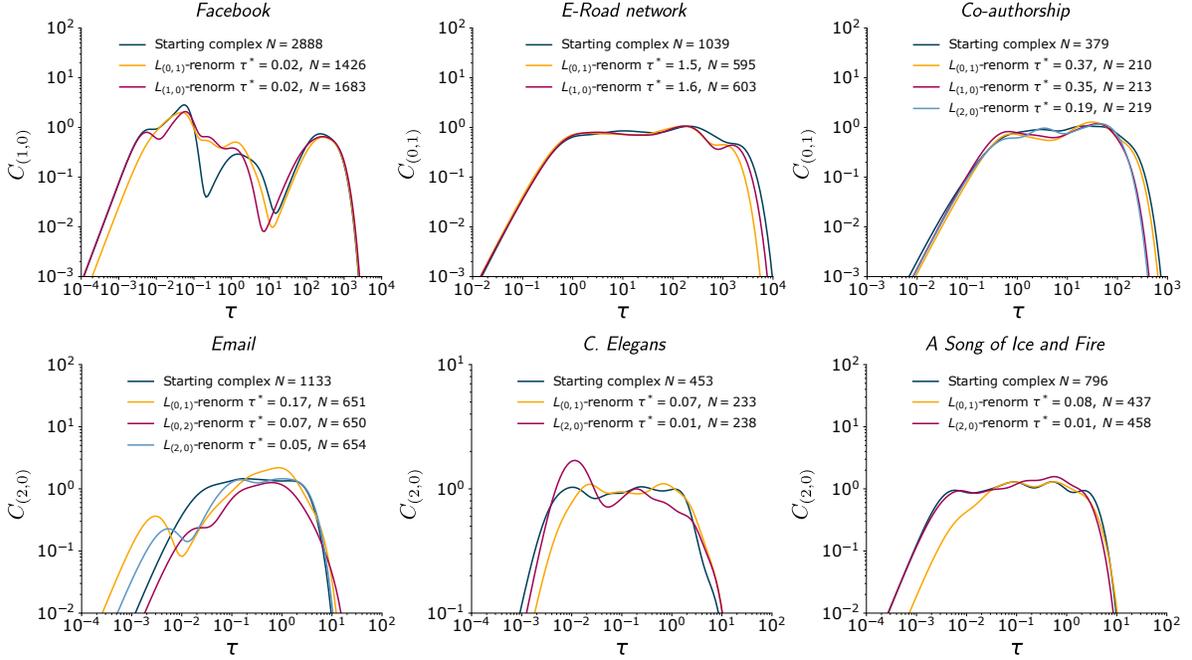

FIG. S4. Entropic susceptibility curves – associated to the highest SIP – of the real data clique-complexes before and after renormalization with different cross-order Laplacians.

### B. Comparison with the Bayesian hypergraph reconstruction method

Up to now we built simplicial complexes from networks by considering their associated clique complexes $\Delta_c$, i.e. by filling all of their cliques. One may argue, however, that not all cliques of nodes correspond to actual higher-order interactions. Our SIP measures then effectively refer to relations between cliques and not higher-order interactions.

To check the dependence of our results with respect to this assumption, we considered simplicial complexes $\Delta_R$ obtained from the networks with the Bayesian hypergraph reconstruction method of Ref. [12]. Note that with this probabilistic method, we do not fill all the cliques of the network, but only the ones for which there is sufficient statistical. We thus obtain a subcomplex of the clique complex $\Delta_R \subseteq \Delta_C$. In Figure S5a we show the scale-invariance parameters of $\Delta_R$ for each dataset (in color), comparing them with the ones obtained on the clique complexes $\Delta_C$ (in gray). Using this more refined reconstruction method, the parameters seem to have smaller values than with the clique complex. In particular, we see that the high values of higher-order SIP in $\Delta_C$ drop to negligible values in $\Delta_R$, meaning that the scale-invariance of the structural organization of cliques is lost.

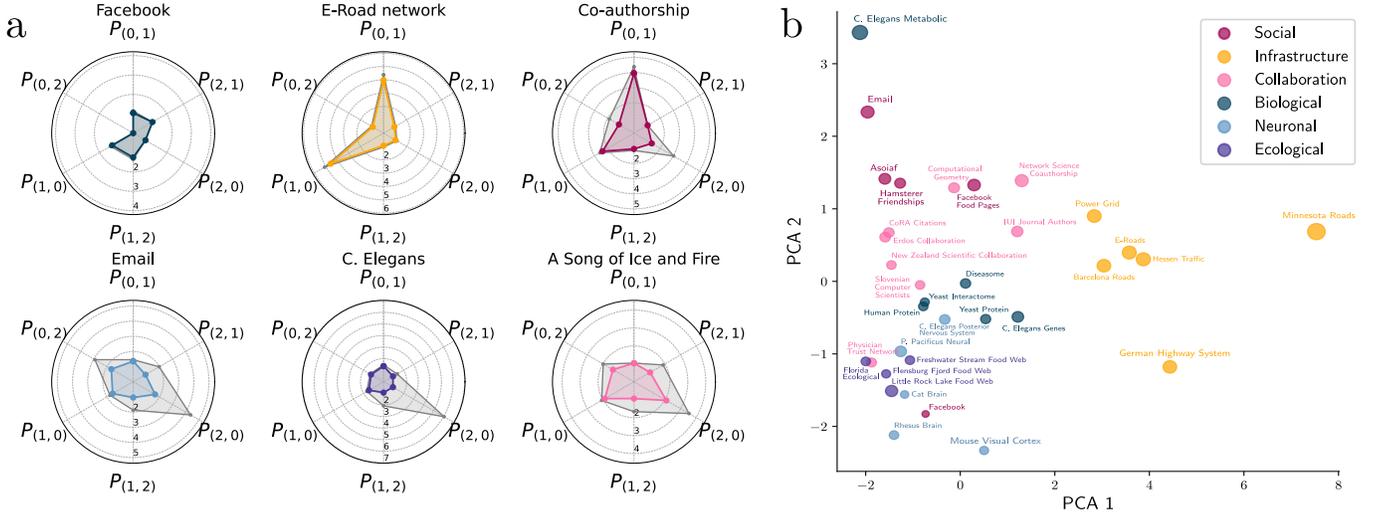

FIG. S5. **a.** Scale-invariance parameters for the 2nd order simplicial complexes obtained with the hypergraph reconstruction method (in color) and the clique complexes (in gray). **b.** A more detailed version of the Figure 6 in the main text. The projections of the 34 datasets onto the first two PCA components is shown, together with their name. The color of each point represents the type of the associated network, while the size is proportional to the value of its highest SIP.

## C.  scale-invariance parameters of real datasets

As we mentioned in the main text, we then considered a larger set of datasets taken from different network archives: *KONECT* [11], *ICON* [13] and *The Network Data Repository* [14]. The networks, whose details are found in Table I, were chosen to belong to the following classes:

- *Infrastructure* — road networks and power grids;

- *Collaboration* — scientific collaboration networks;

- *Social* — email networks, real and fictional social networks;

- *Biological* — gene interaction networks and protein-protein interaction networks;

- *Connectome* — brain networks;

- *Ecological* — species food webs in different ecological environments.

For each one of them, we compute the scale-invariance parameters up to the second order, that is $P_{(0,1)}$, $P_{(0,2)}$, $P_{(1,0)}$, $P_{(1,2)}$, $P_{(2,0)}$ and $P_{(2,1)}$ and use the resulting values as coordinates to embed it as a point in $\mathbb{R}^6$. The resulting point cloud is then projected onto its first two principal components (81.2% of total variance explained) allowing us to get Figure S5b. As we can see, networks of similar type tend to correspond to closer points. Infrastructure networks in yellow are all on the right, the collaboration and social networks, which are necessarily similar in nature, occupy the left side. In the bottom side, we see biological networks, connectomes and ecological food webs, which, while they appear more mixed, they are all of biological origin.

| Name | Type | Repository | Source |
|---|---|---|---|
| *Facebook* | Social | KONECT | [15] |
| *E-Road* | Infrastructure | KONECT | [16] |
| *Network Science Coauthorship* | Collaboration | KONECT | [17] |
| *Email* | Social | KONECT | [18] |
| *C. Elegans Metabolic* | Biological | KONECT | [19–21] |
| *A Song of Ice and Fire* | Social | KONECT | [22] |
| *Human Protein* | Biological | KONECT | [23] |
| *Power Grid* | Infrastructure | KONECT | [24] |
| *Yeast Protein* | Biological | KONECT | [25–28] |
| *Hamsterer Friendships* | Social | KONECT | [11] |
| *German Highway System* | Infrastructure | ICON | [29] |
| *Cat Brain* | Connectome | ICON | [30] |
| *Rhesus Brain* | Connectome | ICON | [31] |
| *Mouse Visual Cortex* | Connectome | ICON | [32] |
| *C. Elegans Posterior Nervous System* | Connectome | ICON | [33] |
| *IUI Journal Authors* | Collaboration | ICON | [34] |
| *CoRA Citations* | Collaboration | ICON | [35] |
| *Flensburg Fjord Food Web* | Ecological | ICON | [36] |
| *Hessen Traffic* | Infrastructure | ICON | [37] |
| *Minnesota Roads* | Infrastructure | Net. Data. Rep. | [38] |
| *Yeast Interactome* | Biological | ICON | [39] |
| *Barcelona Roads* | Infrastructure | ICON | [40] |
| *Freshwater Stream Food Web* | Ecological | ICON | [41] |
| *Little Rock Lake Food Web* | Ecological | ICON | [42] |
| *P. Pacificus Neural* | Connectome | ICON | [43] |
| *Slovenian Computer Scientists* | Collaboration | ICON | [44] |
| *New Zealand Scientific Collaboration* | Collaboration | ICON | [45] |
| *Computational Geometry* | Collaboration | ICON | [46] |
| *Physician Trust Network* | Collaboration | ICON | [47] |
| *Erdos Collaboration* | Collaboration | ICON | [48] |
| *Facebook Food Pages* | Social | Net. Data. Rep. | [49] |
| *Diseasome* | Biological | Net. Data. Rep. | [50] |
| *C. Elegans Genes* | Biological | Net. Data. Rep. | [51] |
| *Florida Ecological* | Ecological | Net. Data. Rep. | [52, 53] |

TABLE I. Datasets used for Figure S5b



\end{document}